# Supercontinuum Generation in Heavy-Metal Oxide Glass Based Suspended-Core Photonic Crystal Fibers

A. N. Ghosh,[1] M. Klimczak,[2,3] R. Buczynski,[2,3] J. M. Dudley,[1] and T. Sylvestre [1, *]

[1]*Institut FEMTO-ST, CNRS, UMR 6174, Université Bourgogne Franche-Comté, Besançon, France*
[2]*Institute of Electronic Materials Technology, Wolczynska 133, 01-919 Warsaw, Poland*
[3]*University of Warsaw, Faculty of Physics, Pasteura 5, 02-093 Warsaw, Poland*
[*]*Corresponding author:* thibaut.sylvestre@univ-fcomte.fr

**We investigate supercontinuum generation in several suspended-core soft-glass photonic crystal fibers pumped by an optical parametric oscillator tunable around 1550 nm. The fibers were drawn from lead-bismuth-gallium-cadmium-oxide glass (PBG-81) with a wide transmission window from 0.5-2.7 µm and a high nonlinear refractive index up to $4.3 \cdot 10^{-19}$ m$^2$/W. They have been specifically designed with a microscale suspended hexagonal core for efficient supercontinuum generation around 1550 nm. We experimentally demonstrate two supercontinuum spectra spanning from 1.07-2.31 µm and 0.89-2.46 µm by pumping two PCFs in both normal and anomalous dispersion regimes, respectively. We also numerically model the group velocity dispersion curves for these fibers from their scanning electron microscope images. Results are in good agreement with numerical simulations based on the generalized nonlinear Schrödinger equation including the pump frequency chirp.**

## 1. INTRODUCTION

Broadband, compact and cost-effective supercontinuum (SC) sources in the infrared (IR) are very attractive for sensing and spectroscopy applications [1-3] because this wavelength range contains the ground tones of many molecules, yielding unique "fingerprints" of their chemical composition [4]. However, the state-of-the-art mid-IR SC systems are still in their infancy, thus motivating further efforts in fiber drawing and in soft-glass chemistry to remove absorption bands while enhancing both transmission and nonlinearity. Soft glasses such as chalcogenide ($AS_2S_3$) [4-6], tellurite [7,8], and ZBLAN ($ZrF_4$ –$BaF_2$ –$LaF_3$ –$AlF_3$ –NaF) [9-11] have been widely used for drawing highly nonlinear fibers, and experiments have shown efficient SC generation in the mid-IR up to 15 µm [4-5] and up to 11 µm using all-fiber systems [12-13]. Another approach for mid-IR SC generation is gas-filled hollow-core photonic crystal fiber, which provides low-loss guidance and much higher effective nonlinearity than capillary based system [14,15]. Record output powers up to 3.2 µm has also been recently achieved in highly Germania doped fibers [16].

An alternative to ZBLAN fibers that has recently received attention is small-core heavy metal oxide photonic crystal fibers (PCFs) [17]. They possess a number of optical and mechanical properties making them also attractive for SC generation, despite their limited transmission window from 550-2800 nm, compared to other soft glasses. The nonlinear index $n_2$ of this glass is one order of magnitude larger ($4.3 \times 10^{-19}$ m$^2$/W) than ZBLAN and silica glasses [17]. In particular, suspended core photonic crystal fibers (SC-PCF) can be readily drawn in which the microstructures enabled have tight optical confinement and small effective mode area in the order of few µm$^2$ [18,19]. This results in high nonlinear coefficient with a drawback of lower coupling efficiency to the fiber due to small core diameter. In the recent years, suspended-core fibers received less attention in the context of supercontinuum generation than regular lattice PCFs, although the existing literature reports provide evidence on the versatility of this approach, including choice of glass platform (silica/soft glass), spectral coverage (NIR/MIR) and dispersion regime of SC generation (solitonic/all-normal) [20-24].

Here, we report SC generation in lead-bismuth-gallium-cadmium-oxide glass (PBG-81) based suspended-core PCF pumped by 200 fs pulses (delivered by a Ti-Sapphire laser pumped OPO system) around 1550 nm.
Using two SC-PCFs (core diameter = 4.5 µm and 3.5 µm) with respective lengths of 60 cm and 180 cm, we demonstrate SC generation spanning 1.07-2.31 µm with an output mean power of 53 mW and 0.89-2.46 µm with an output mean power of 32 mW, respectively by pumping at 1550 nm and 1580 nm. We identify a number of nonlinear phenomena such as spectral broadening due to self-phase modulation, soliton generation, and Raman soliton self-frequency shift in the fiber at the pumping wavelengths. The experimental spectra were compared with numerical simulations of the generalized nonlinear Schrödinger equation including the group velocity dispersion computed from the scanning electron microscope (SEM) images of the fiber samples.

## 2. HEAVY-METAL-OXIDE SUSPENDED-CORE PHOTONIC CRYSTAL FIBERS

The suspended-core photonic crystal fibers were drawn from lead-bismuth-gallium-cadmium-oxide (PBG81) glass. Its chemical composition contains the oxides as follows: PbO: 39.17, $Bi_2O_3$: 27.26, $Ga_2O_3$: 14.26, $SiO_2$: 14.06 and CdO: 5.26. The nonlinear refractive index of PBG81 glass was measured as $n_2 = 43 \times 10^{-20}$ $m^2$/W (Z-scan method at 1240 nm, bulk sample) [25,26]. The attenuation spectrum of this PBG-81 bulk glass is shown in Fig. 1. This soft glass has a transmission window from 550-2800 nm with typical attenuation of around 2.5 dB/m within 1000-2000 nm. Above 2800 nm, the particular glass melt shows onset of strong attenuation assigned to OH absorption. This particular glass melt has been subject only to standard bubbling of the glass melt with dry air (<1000 ppm of OH). As can be seen in Fig. 1, the attenuation curve does not feature a local peak around 1380 nm. This is the main difference between PBG-81 and previously-used PBG-08 glass [17].

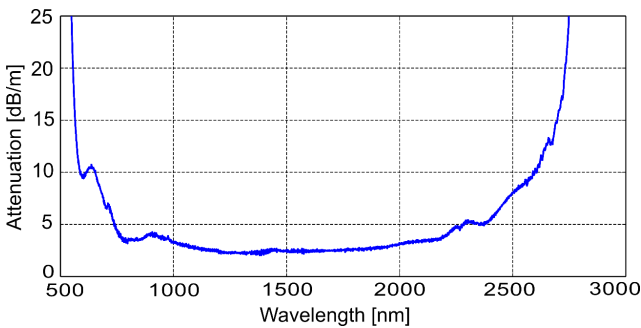

**Fig. 1.** Attenuation spectrum of bulk lead-bismuth-gallium-cadmium-oxide glass used for drawing SCPCFs

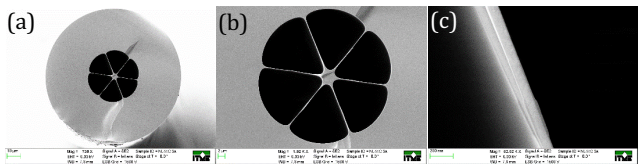

**Fig. 2.** Cross-section SEM images of a suspended core PCF: (a) Image of the fiber with outer diameter of 99.5 μm. (b) Expanded view of the microstructure region (core diameter of 3.95 μm). (c) One of the 6 struts supporting the core with thickness of 89.7 nm. The microstructure of one of the five PCFs is shown in Figs. 2(a-c). The PCF's outer and core diameters are 99.5 and 3.95 μm, and there are 6 struts or bridges supporting the hexagonal core. These thin struts result in a negligible modal confinement loss for the fundamental mode by ensuring a high degree of isolation in the core area. We fabricated five different fibers with a microscale suspended hexagonal core with dimensions chosen to yield zero-dispersion wavelengths (ZDWs) covering the range 1519-1653 nm.

**Table 1. Geometric parameters of SC-PCFs used for dispersion calculation and supercontinuum generation**

| No. | Label | Core diameter (μm) | Strut thickness (nm) | ZDW (nm) |
|---|---|---|---|---|
| 1 | NL44C1a | 4.5 | 81 | 1653 |
| 2 | NL44C2a | 4.3 | 99 | 1593 |
| 3 | NL44C3a | 3.95 | 90 | 1574 |
| 4 | NL44C4b | 3.8 | 40 | 1557 |
| 5 | NL44C5c | 3.51 | 47 | 1519 |

The opto-geometric parameters of the five fibers are given in Table 1. From SEM images (see for example Fig. 2(b)), we calculated the effective refractive index of the fundamental mode and the corresponding group-velocity dispersion (GVD) using COMSOL software. We used a standard Sellmeier equation and the coefficients used in the simulation were: $B_1$= 2.30350920, $B_2$= 0.21430548, $B_3$= 1.73310331, $C_1$= 0.02084623 $\mu m^2$, $C_2$ = 0.08262994 $\mu m^2$, and $C_3$ = 183.5615768 $\mu m^2$. The numerical results are shown in Figs. 3(a) and 3(b) for our five fiber samples.

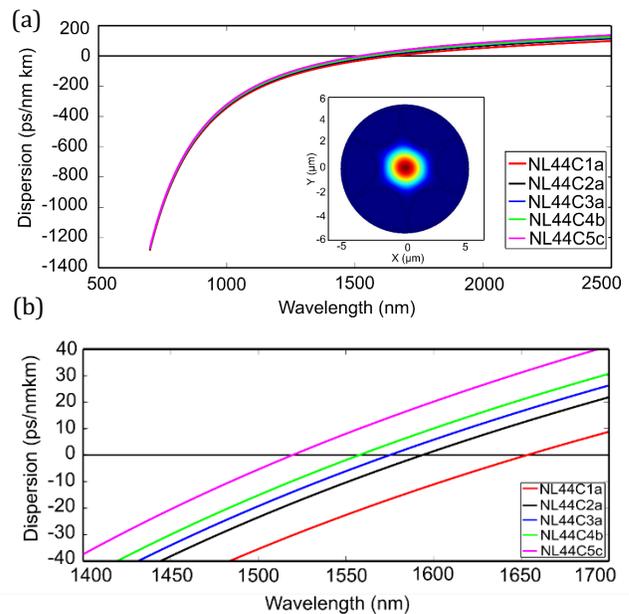

**Fig. 3.** Dispersion characteristics of the SC-PCFs simulated for the fundamental mode from SEM images: (a) Group velocity dispersion of 5 SC-PCF samples (Inset: optical power density of the fundamental mode inside the core of fiber sample NL44C2a). (b) Dispersion curve of the PCFs showing zero-dispersion wavelength (ZDW).

Figure 3(a) shows the group velocity dispersion curve of our five SC-PCF samples covering almost the entire transmission window of the fiber and the inset image represents the optical power density of the fundamental mode inside the core of fiber sample NL44C2a. As can be seen in Fig. 3(b), the zero-dispersion wavelength shifts from 1519 nm to 1653 nm as the core diameter increases from 3.51 μm to 4.5 μm.

## 3. EXPERIMENTAL SETUP

The experimental setup for generating and measuring supercontinuum spectra is shown in Fig. 4. As a pump laser we used a 200-fs Optical Parametric Oscillator (Chameleon Compact OPO-Vis), pumped by a Ti-Sapphire mode-locked laser (Chameleon Ultra II) at 80 MHz repetition rate. The OPO signal is tunable in the range 1-1.6 µm with mean output powers 1 W - 230 mW, and the idler from 1.7 µm to 4 µm with powers from 250 mW to 50 mW. The signal from the OPO was injected into the SC-PCF through a focusing objective and the coupling power was controlled via a variable attenuator. An IR camera and imaging system (another focusing objective) were combined with wavelength filters to study the mode structure and guidance inside the fiber core. The generated SC light was measured using an IR spectrometer (Ocean Optics NIRQuest512-2.5) with sensitivity in the wavelength range of 900-2500 nm. The result of the IR imaging showed that the SC was generated in the fundamental mode over the full wavelength range and no evidence for the presence of multimode operation could be observed.

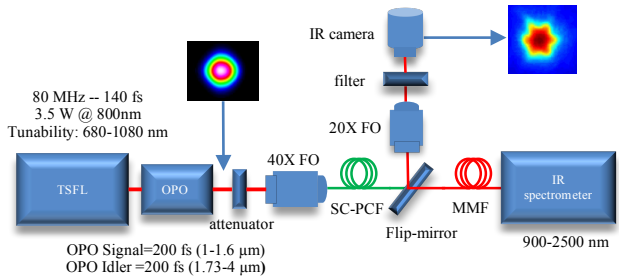

**Fig. 4.** Scheme of the experimental setup for generating and measuring supercontinuum infrared light. TSFL, Ti-Saphire femtosecond laser; FO, focusing objective; SC-PCF, suspended core photonic crystal fiber; MMF, multimode fiber [inset-left: beam profile at OPO output, right: IR image at SC-PCF output].

## 4. RESULTS AND DISCUSSIONS

Figures 5(a)-(d) shows the variation in the measured SC spectra for different fiber output powers as indicated for two SC-PCF samples NL44C2a (length = 60 cm, ZDW = 1593 nm) and NL44C5c (length = 180 cm, ZDW = 1519 nm) using signal wavelengths of 1550 nm (in the normal dispersion regime) and 1580 nm (in the anomalous dispersion regime), respectively.

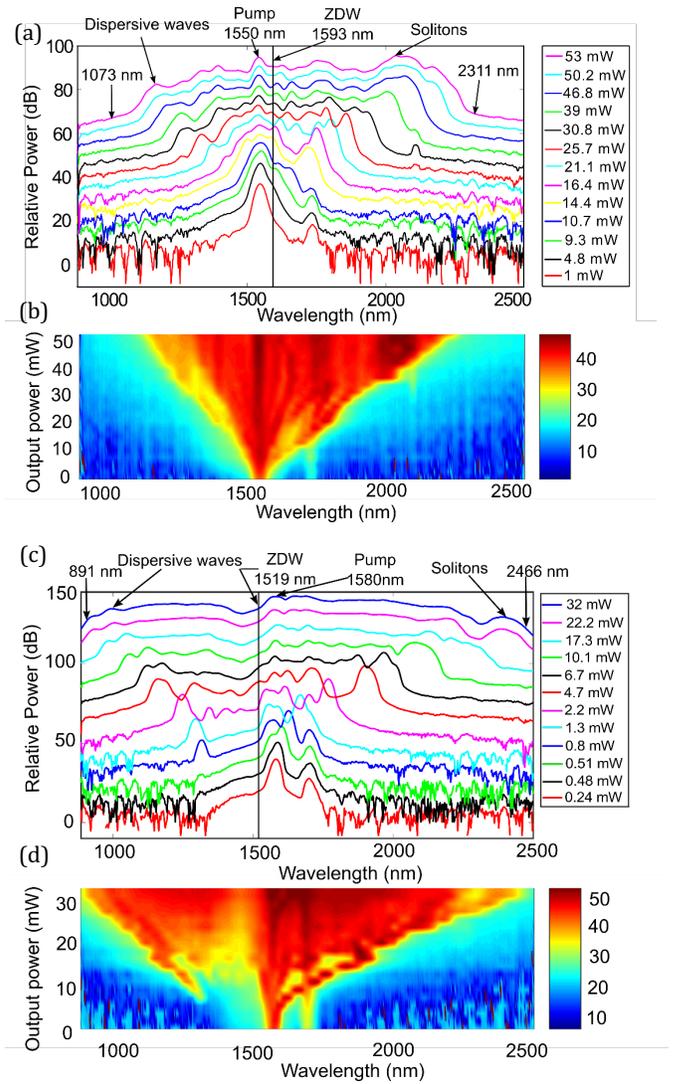

**Fig. 5.** Supercontinuum spectra generated in two SC-PCF samples with 200 fs pulses pumped at 1550 nm and 1580 nm, respectively, as a function of mean output power. (a) and (c) Generation of SC spectra through spectral broadening, soliton ejection, and dispersive wave generation in two fiber samples NL44C2a and NL44C5c, respectively. (b) and (d) Evolution of SC spectrum with the fiber output power for SC-PCF samples NL44C2a and NL44C5c.

The output spectra recorded for these two fiber samples show strong spectral broadening around the pump wavelength due to self-phase modulation at low input power. Note that in addition to the peak contributed by pump wavelength we have an additional peak due to the residual OPO idler, which otherwise does not contribute to SC generation. As the power increases, we see red-shifted solitons beyond 2 µm and simultaneous generation of up-shifted dispersive waves down to 1 µm. At maximum coupling power, we obtain SC spectra of 1238 nm and 1575 nm bandwidths (at the -25 dB and -20 dB level, respectively) for fiber samples NL44C2a and NL44C5c, respectively.

A SC-PCF sample NL44C4b (length = 183 cm, ZDW = 1557 nm) is also pumped at 1740 nm in the strong anomalous dispersion regime using the idler OPO output. The variation in the measured SC spectra for different output powers for this fiber is shown in Fig. 6. As we pumped at a wavelength far away from the zero-dispersion wavelength of this fiber, we obtained three individual high power solitons beyond 2 μm and dispersive wave down to 1 μm instead of a smooth supercontinuum. At maximum power, the bandwidth of this broad spectrum is 1495 nm at the -25 dB level.

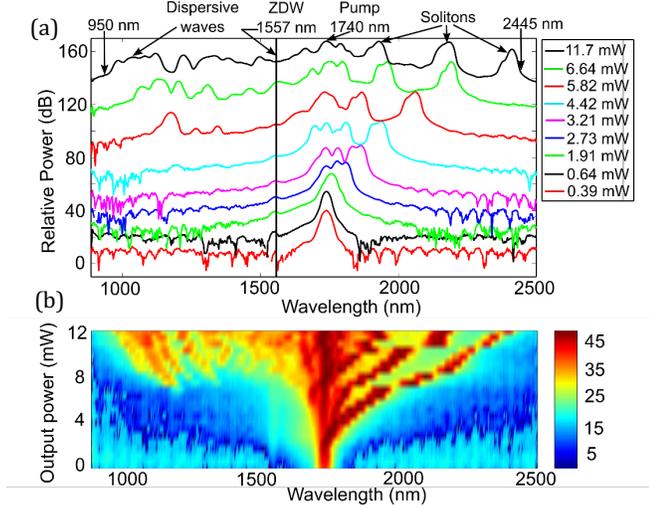

**Fig. 6.** Spectra generated in SC-PCF sample NL44C4b with 200 fs pulses pumped at 1740 nm. (a) Generation of broad spectra through spectral broadening, soliton ejection, and dispersive wave generation. (b) Evolution of spectrum with the fiber output power.

## 5. SUPERCONTINUUM SIMULATION

Nonlinear pulse propagation and supercontinuum generation was modelled using generalized nonlinear Schrödinger equation (GNLSE) [27,28]

$$\frac{\partial A}{\partial z} + \frac{\alpha}{2} A - \sum_{k \geq 2} \frac{i^{k+1}}{k!} \beta_k \frac{\partial^k A}{\partial T^k} = i\gamma \left(1 + i\tau_{shock} \frac{\partial}{\partial T}\right) \times \left(A(z,T) \int_{-\infty}^{+\infty} R(T') |A(z, T - T')|^2 dT'\right), \quad (1)$$

Here the second term in the left-hand side of Eq. (1) accounts for linear loss with loss coefficient $\alpha$, and the third term represents the dispersion with dispersion coefficient $\beta_k$ associated with Taylor series expansion of the propagation constant $\beta(\omega)$ about central frequency $\omega_0$. In the simulation, we introduced a linear loss of 5 dB/m (measured using cut-back method). We also used the dispersion parameters (up to fourth order) computed from SEM image of SC-PCF samples. The right-hand side (RHS) of Eq. (1) models the nonlinear effects in the fiber: nonlinear coefficient $\gamma = 2\pi n_2/\lambda_0 A_{eff}$, where nonlinear refractive index $n_2$ was $4.3 \times 10^{-19}$ m$^2$/W and effective mode area $A_{eff}$ was calculated from SEM image of fiber sample at pump wavelength. The time derivative term in RHS of Eq. (1) models the effects such as self-steepening and optical shock formation with $\tau_{shock} = 1/\omega_0$. The term $R(T) = (1 - f_R)\delta(t) + f_R h_R(t)$ represents the response function which includes both instantaneous electronic and delayed Raman contribution with $f_R$=0.05. The value of $f_R$ was measured from Raman scattering spectra of bulk glass and Raman contribution to Kerr nonlinearity $h_R(t)$ was fitted from nonlinear simulation for a different type of fiber made from the same glass. The input pulse was considered to be Gaussian. Frequency-resolved optical gating measurements of our input pulses (MeasPhotonics FROGscan) showed the presence of a linear chirp (parabolic phase) which was also included in the simulation initial conditions. In the simulation for SC-PCF samples NL44C2a and NL44C5c, we used fiber lengths of 60 cm and 180 cm, and pump pulse duration of 225 fs and 220 fs, respectively. The peak power at the input of fiber was considered to be 25 kW for the both cases. Figures 7(a)-(d) shows the evolution of numerically generated supercontinuum spectra along fiber length in spectral and time domain for the two previously mentioned fiber samples pumped at 1550 nm and 1580 nm. We only show the first beginning of pulse propagation to highlight the main nonlinear effects in SC generation. In Figs. 7(a) & (c), strong spectral broadening can be seen in the beginning of fiber due to self-phase modulation. With increase in length, we can see soliton ejection and generation of dispersive waves. The soliton distribution as a function of time are shown in Figs. 7(b) & (d). The comparisons between simulation and experiment in the 1550 nm and 1580 nm pump case are shown in Fig. 8. Both the simulated and experimental SC spectra almost agree with each other in terms of SC bandwidth but they are quite different in terms of structure of the spectrum as can be seen from Fig. 8(a) for SC-PCF sample NL44C2a. We speculate that the observed discrepancy is accountable to some assumptions in the model. First, we did not take into account the wavelength-dependent losses over the full SC span and the higher-order optical modes. Second, the dispersion profile was numerically estimated from the cross-section image of the fiber structures. It is reasonable that small change of the slope of normal dispersion at wavelengths blue-shifted from the pump wavelength, would result in different dispersive wave phase matching conditions, which would most reasonable explain the short-wavelength discrepancies between the theoretical and measured spectra in Fig. 8. The relatively long fiber samples used in the experiment (for soft glass fibers and femtosecond pump pulses), would suggest significant contribution of the OH absorption loss, however, no such features are observed at around 1380 nm in spectra shown in Fig. 8. Bend loss of any higher-order mode-contained blue-shifted parts of the supercontinuum spectra could also be related the observed discrepancies, but we did not observe such loss behavior for any reasonable bend radii in these fibers during the measurements.

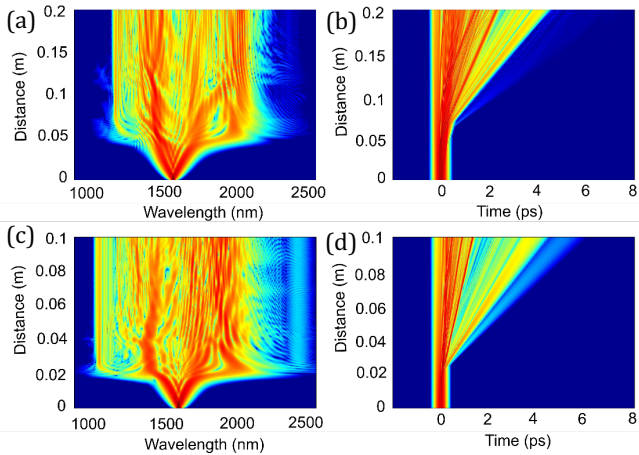

**Fig. 7.** (a) & (c) Numerically generated evolution of SC spectra along the fiber length for NL44C2a & NL44C5c PCFs with 1550nm and 1580nm pumping, respectively. (b) & (d) Corresponding temporal distribution along the fiber length for NL44C2a & NL44C5c PCFs.

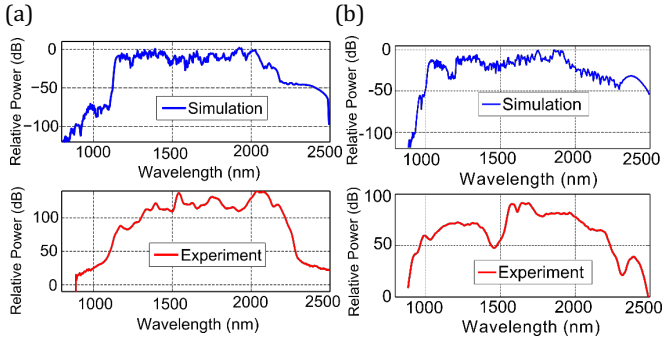

**Fig. 8.** Comparison between numerically and experimentally generated SC spectra for 1550 nm and 1580 nm pumping in (a) NL44C2a and (b) NL44C5c PCFs.

## 6. CONCLUSION

Wideband SC generation from 1073 nm - 2311 nm and from 891 nm - 2466 nm was recorded by pumping two suspended core photonic crystal fibers in the both normal and anomalous dispersion regimes, respectively, with femtosecond pulses at 1550 nm and 1580 nm. These results demonstrate the utility of heavy-metal-oxide PCF for SC generation. Numerical simulations using nonlinear Schrödinger equation and the group velocity dispersion computed from the scanning electron microscope (SEM) images of the SC-PCF samples show good qualitative agreement with the experimental one for 4.3 µm and 3.51 µm core diameter SC-PCF samples. It is important to stress that the SC spectra obtained using PBG-81 glass suspended-core optical fibers are limited by their transmission windows up to 2.8 µm, which may limit their potential use for e.g., infrared spectroscopy. They are actually comparable to those can be obtained with highly nonlinear germanium-doped optical fibers [26].

## 7. FUNDING

This work was funded by the European Union's Horizon 2020 research and innovation program under grant agreement no. 722380, the I-SITE UBFC (ANR-15-IDEX-0003), the Labex Action (ANR-11-LABX-0001), the SONATA project UMO-2013/11/D/ST7/03156 awarded by National Science Centre in Poland, and First TEAM/2016-1/1 project awarded by the Foundation for Polish Science. The authors thank Damien Bigourd, Gil Fanjoux and Alexis Mosset for technical supports.

**FULL REFERENCES**